\documentclass{article}

\newcommand{\IGN}[1]{{}}

\usepackage{epic}
\usepackage{eepic}   
\usepackage{epsfig}
\newcommand{\emline}[6]{\drawline(#1,#2)(#4,#5)}

\usepackage{here}

\usepackage{times}
\usepackage{multicol}

\begin{document}
%
%
%
%
\title{
Winner-Relaxing and Winner-Enhancing Kohonen Maps: 
\\
Maximal Mutual Information from Enhancing the Winner
}
%
%
\author{Jens Christian Claussen
\\
Institut f\"ur Theoretische Physik und Astrophysik\\
Christian-Albrechts-Universit\"at zu Kiel,
\\
Leibnizstr. 15, 24098 Kiel, Germany
}

%
%


\date{\small July 9, 2002. Published in: 
\underline{Complexity (Wiley/VCH) {\bf 8} (4),
15-22 (2003)}}

\maketitle              

\begin{abstract}
The magnification behaviour of a
generalized family of self-organizing 
feature maps,
the Winner Relaxing  and  Winner Enhancing 
Kohonen algorithms
is analyzed by the magnification law in the
one-dimensional case, which can be obtained analytically.
The Winner-Enhancing case allows
to acheive a magnification exponent of one
and therefore provides optimal mapping in the sense of
information theory. 
A numerical verification of the magnification law is
included, 
and the ordering behaviour is analyzed.
Compared to the original Self-Organizing Map and
some other approaches, the generalized 
Winner Enforcing Algorithm requires minimal extra computations
per learning step and is conveniently easy to implement.
\end{abstract}

\normalsize \rm


\begin{multicols}{2}

The Self-Organizing Map defined by Kohonen in 1982
\cite{kohonen82}
startet a class of highly successful neural network models,
since it has shown both relevance for
modeling of biological networks and
engineering of artificial neurocomputing
including data analysis
and self-organized dimension reduction of complex
structured and high-dimensional input spaces.

The qualitative biological inspiration
comes from cortical receptor fields as the 
receptor fields of the retina, and the 
receptor field of the skin. 
For both, neighboured sensory input 
always will be represented by cortical 
activity that is also of neighboured 
location in the neural tissue.
Such topology-preserving 
{\sl neural maps}
are known quite a long,
and apart from error-tolerant computation,
two striking properties are known:
First, there is high plasticity, 
e.g. when a finger is cut off, the 
neurons now lacking sensory information
begin to specialize themselves for 
the adjacent fingers.
Second, as the complete structure 
(of all synaptic weights even of parts of the brain)
is far too complex to be coded genetically, 
the detailed structure has to emerge from
a self-organizing process,
obviously stochastically driven by the 
sensory information.

While the qualititative structure 
of biological maps can successfully be
modeled even with the simple Kohonen map, 
there are many variants and modifications
leading to qualitatively similar
self-organizing topology-preserving maps.
The approaches to define and discuss these
are as different as the
resulting algorithms, but can be categorized
roughly as follows.
(a) Derivation from first principles,
as energy or cost functions, mutual information,
averaged representation error, distortion
measures and every combination of these.
(b) Argumentation from structure,
from a realistic biological model
to discussion of optimal technical implementations.
(c) Extraction of measurable quantitative properties,
as magnification laws, properties of fluctuation,
ordering, adaptation, error-tolerance, and
spatial frequency of singularities upon dimension
reduction, as in the retina.
Finally (d) Restriction to the simplest possible models,
which is merely a physicists point of view.

As there are different aims of using feature maps,
these may naturally lead to different viewpoints.
For some technical applications, it may be convenient to use any 
kind of vector quantization, e.~g. the Kohonen model {\it wihout}
any neighborhood interactions, and to apply some sorting algorithm
to set up topological order afterwards. In the brain, hovever, it is
assumed that the topological structure is set up by self-organization,
therefore here we focus on self-organizing maps only.

Compared to the Elastic Net Algorithm of Durbin and Willshaw 
\cite{durbinwillshaw}
and the Linsker Algorithm \cite{linsker89}
which are performing gradient descent in a certain
energy landscape, the Kohonen algorithm seems to have no energy
function.
Although the learning process can be described in terms of
a Fokker-Planck equation \cite{ritter88}, the expectation value of
the learning step is a nonconservative force \cite{obermayer92}
driving the process so that it has no associated energy function.
Furthermore, the relationships of the
Kohonen model to both alternative models and general principles
are still an open field \cite{kohonen91}

\clearpage

In this paper we analyze an algorithm which is a  gradient
system (and therefore has an extremal principle)
in terms of the principle of maximizing mutual information.
As one has to assume the existence of a functional that is
optimized by evolution, the question of extremal principles is central in
theoretical brain research.
Mutual information is assumed to play an essential role
in any energy landscape that may describe the evolution of the
brain.
To measure optimality in sense of information theory
we consider the magnification law for onedimensional maps.
The magnification factor is defined as the number of
synaptic weight vectors (respectively neurons) per unit volume
of input space.
Maps of maximal mutual information show a power law with exponent 1,
but the algorithm given by Linsker \cite{linsker89}
requires a complicated learning rule, whereas an exponent 0
corresponds to no adaptation to the stimuli at all.
The exponent 1 is equivalent to that all neurons have the same
firing probability.

The leading question is if there are models that are suitable to
describe biological maps and show a sufficient magnification
behavior.
The exponents for the Kohonen \cite{ritter86}
and the Linsker algorithm are known quite a long.
Also the Elastic Net also shows a universal 
(i. e. depending only on the local stimulus density)
magnification law which however is 
not a power law \cite{claussenicann},
and for serial presentation does not allow for both 
stability and infomax mapping.

 With the generalized Winner Relaxing Kohonen algorithm
howver,
 by inverting the `relaxing' term
an exponent 1 can be acheived, with a minimal 
computational extension of the algorithm.

\section{\large The Kohonen Self Organizing Feature Map}

The Kohonen algorithm for Self Organizing Feature Maps
is defined as follows:
Every stimulus $\vec{v}$ of an euclidian input space $V$
is mapped 
to the neuron with the position $\vec{s}$
in the neural layer $R$
which has the weight vector
 $\vec{w}_{\vec{s}}$ 
with the 
minimal distance in input space, corresponding to the
highest neural activity, 
which is called
the `center of excitation' or `winner'
(Fig. 
1).

\end{multicols}

\mbox{}\\ 
\begin{center}
\unitlength=.9mm
\linethickness{0.4pt}
\begin{picture}(175.00,75.00) 
\put(90.00,70.00){\line(1,0){70.00}}
\put(90.00,60.00){\line(1,0){70.00}}
\put(90.00,50.00){\line(1,0){70.00}}
\put(90.00,40.00){\line(1,0){70.00}}
\put(90.00,30.00){\line(1,0){70.00}}
\put(90.00,20.00){\line(1,0){70.00}}
\put(90.00,70.00){\line(0,-1){50.00}}
\put(100.00,70.00){\line(0,-1){50.00}}
\put(110.00,70.00){\line(0,-1){50.00}}
\put(120.00,70.00){\line(0,-1){50.00}}
\put(130.00,70.00){\line(0,-1){50.00}}
\put(140.00,70.00){\line(0,-1){50.00}}
\put(150.00,70.00){\line(0,-1){50.00}}
\put(160.00,70.00){\line(0,-1){50.00}}
\put(120.00,50.00){\circle{2.00}}
\put(120.00,50.00){\circle{5.00}}
\put(120.00,50.00){\circle{7.00}}
\put(120.00,60.00){\circle*{4.00}}
\put(130.00,50.00){\circle*{4.00}}
\put(120.00,40.00){\circle*{4.00}}
\put(110.00,50.00){\circle*{4.00}}
\put(110.00,60.00){\circle*{2.83}}
\put(130.00,60.00){\circle*{2.83}}
\put(130.00,40.00){\circle*{2.83}}
\put(110.00,40.00){\circle*{2.83}}
\put(90.00,70.00){\circle*{2.00}}
\put(90.00,60.00){\circle*{2.00}}
\put(90.00,50.00){\circle*{2.00}}
\put(90.00,40.00){\circle*{2.00}}
\put(90.00,30.00){\circle*{2.00}}
\put(90.00,20.00){\circle*{2.00}}
\put(100.00,70.00){\circle*{2.00}}
\put(100.00,60.00){\circle*{2.00}}
\put(100.00,50.00){\circle*{2.00}}
\put(100.00,40.00){\circle*{2.00}}
\put(100.00,30.00){\circle*{2.00}}
\put(100.00,20.00){\circle*{2.00}}
\put(110.00,70.00){\circle*{2.00}}
\put(110.00,30.00){\circle*{2.00}}
\put(110.00,20.00){\circle*{2.00}}
\put(120.00,70.00){\circle*{2.00}}
\put(120.00,30.00){\circle*{2.00}}
\put(120.00,20.00){\circle*{2.00}}
\put(130.00,70.00){\circle*{2.00}}
\put(130.00,30.00){\circle*{2.00}}
\put(130.00,20.00){\circle*{2.00}}
\put(140.00,70.00){\circle*{2.00}}
\put(140.00,60.00){\circle*{2.00}}
\put(140.00,50.00){\circle*{2.00}}
\put(140.00,40.00){\circle*{2.00}}
\put(140.00,30.00){\circle*{2.00}}
\put(140.00,20.00){\circle*{2.00}}
\put(150.00,70.00){\circle*{2.00}}
\put(150.00,60.00){\circle*{2.00}}
\put(150.00,50.00){\circle*{2.00}}
\put(150.00,40.00){\circle*{2.00}}
\put(150.00,30.00){\circle*{2.00}}
\put(150.00,20.00){\circle*{2.00}}
\put(160.00,70.00){\circle*{2.00}}
\put(160.00,60.00){\circle*{2.00}}
\put(160.00,50.00){\circle*{2.00}}
\put(160.00,40.00){\circle*{2.00}}
\put(160.00,30.00){\circle*{2.00}}
\put(160.00,20.00){\circle*{2.00}}
\put(120.00,50.00){\line(-4,-1){79.00}}
\put(41.00,30.00){\vector(-1,1){6.00}}
\put(41.00,30.00){\circle*{2.00}}
\put(123.00,47.00){\makebox(0,0)[lt]{\large$\vec{s}$}}
\put(58.00,73.00){\makebox(0,0)[cc]{{ }}} 
\put(21.00,52.00){\makebox(0,0)[cc]{\large$\vec{v}$}}
\put(43.00,28.00){\makebox(0,0)[lt]{\large$\vec{w}_{\vec{s}}$}}
\put(37.00,38.00)
    {\makebox(0,0)[lb]{\large$\bigtriangleup\vec{w}_{\vec{s}}$}}
\put(59.00,18.00){\makebox(0,0)[cc]{  }} 
\put(136.00,8.00){\makebox(0,0)[cc]{  }} 
\put(24.95,46.00){\line(1,1){3.00}}
\put(27.95,49.00){\line(6,5){6.07}}
\put(34.02,54.02){\line(5,4){5.02}}
\put(39.04,57.99){\line(4,3){3.97}}
\put(43.01,60.92){\line(5,3){5.02}}
\put(48.03,63.85){\line(2,1){3.97}}
\put(52.01,65.73){\line(5,2){6.07}}
\put(58.08,68.24){\line(4,1){5.86}}
\put(63.93,69.71){\line(5,1){6.07}}
\put(70.00,70.96){\line(1,0){6.07}}
\put(76.07,70.96){\line(6,-1){5.86}}
\put(81.92,69.92){\line(3,-1){6.07}}
\put(87.99,67.82){\line(5,-2){5.02}}
\put(93.01,65.73){\line(2,-1){5.02}}
\put(98.03,63.22){\line(2,-1){6.07}}
\put(104.10,60.29){\line(5,-3){3.97}}
\put(108.08,57.99){\line(5,-3){5.02}}
\put(113.10,55.06){\vector(2,-1){1.88}}
\put(24.95,46.00){\circle{2.00}}
\put(24.95,46.00){\circle{5.00}}
\emline{4.98}{56.95}{1}{9.07}{59.08}{2}
\emline{9.07}{59.08}{3}{14.94}{61.04}{4}
\emline{14.94}{61.04}{5}{19.03}{61.92}{6}
\emline{19.03}{61.92}{7}{25.96}{62.99}{8}
\emline{25.96}{62.99}{9}{35.92}{64.06}{10}
\emline{35.92}{64.06}{11}{51.92}{64.06}{12}
\emline{51.92}{64.06}{13}{57.97}{62.99}{14}
\emline{57.97}{62.99}{15}{62.95}{61.92}{16}
\emline{62.95}{61.92}{17}{70.06}{61.04}{18}
\emline{70.06}{61.04}{19}{73.09}{61.04}{20}
\emline{73.09}{61.04}{21}{77.00}{61.92}{22}
\emline{77.00}{61.92}{23}{78.95}{62.99}{24}
\emline{78.95}{62.99}{25}{77.00}{59.97}{26}
\emline{77.00}{59.97}{27}{75.04}{56.06}{28}
\emline{75.04}{56.06}{29}{73.09}{51.08}{30}
\emline{73.09}{51.08}{31}{72.02}{45.92}{32}
\emline{72.02}{45.92}{33}{70.95}{37.03}{34}
\emline{70.95}{37.03}{35}{70.95}{30.98}{36}
\emline{70.95}{30.98}{37}{72.02}{21.91}{38}
\emline{72.02}{21.91}{39}{73.09}{18.00}{40}
\emline{73.09}{18.00}{41}{75.04}{11.96}{42}
\emline{75.04}{11.96}{43}{70.95}{11.07}{44}
\emline{70.95}{11.07}{45}{57.97}{10.00}{46}
\emline{57.97}{10.00}{47}{51.04}{10.00}{48}
\emline{51.04}{10.00}{49}{43.92}{11.07}{50}
\emline{43.92}{11.07}{51}{33.96}{13.02}{52}
\emline{33.96}{13.02}{53}{25.96}{13.91}{54}
\emline{25.96}{13.91}{55}{19.92}{13.91}{56}
\emline{19.92}{13.91}{57}{12.98}{13.02}{58}
\emline{12.98}{13.02}{59}{11.91}{16.94}{60}
\emline{11.91}{16.94}{61}{11.03}{21.91}{62}
\emline{11.03}{21.91}{63}{11.03}{27.07}{64}
\emline{11.03}{27.07}{65}{12.98}{35.07}{66}
\emline{12.98}{35.07}{67}{12.98}{40.05}{68}
\emline{12.98}{40.05}{69}{11.91}{46.99}{70}
\emline{11.91}{46.99}{71}{11.03}{50.01}{72}
\emline{11.03}{50.01}{73}{9.96}{51.97}{74}
\emline{9.96}{51.97}{75}{8.00}{54.99}{76}
\emline{8.00}{54.99}{77}{6.94}{56.06}{78}
\emline{6.94}{56.06}{79}{4.98}{56.95}{80}
\end{picture}
\end{center}
\mbox{}
\\
\noindent
{\sl 
{\bf Fig. 1:}  A Self-Organizing Mapping $\Phi$ from an input space V to a neural layer R.  
Every stimulus $\vec{v}$  gets assigned to a center of excitation $\vec{s}$.  
Weight vectors $\vec{w}_{\vec{s}}$ change 
according to a learning rule 
that defines each map algorithm.
}
\\
\\\\

\begin{multicols}{2}

In the Kohonen model the learning rule for each synaptic
weight vector $\vec{w}_{\vec{r}}$ is given by
        \begin{eqnarray}
\delta
        \vec{ w}_{\vec{r}} =
         \eta \cdot g_{\vec{r} \vec{s}}
         \cdot (\vec{v}-\vec{ w}_{\vec{r}})
        \end{eqnarray}
with $g_{\vec{r}\vec{s}}$ as a gaussian function of euclidian
distance $|\vec{r}-\vec{s}|$ in the neural layer.
The function $g_{\vec{r}\vec{s}}$ describes the topology in the
neural layer. The  parameter $\eta$ determines the speed of
learning and can be adjusted during the learning process.
Topology preservation is enforced by the common
update of all weight vectors whose neuron $\vec{r}$ is adjacent
to the center of excitation $\vec{s}$.

\section{\large 
\mbox{\mbox{}\hspace*{-0.4em}Salesman Travelling to Town and Countryside:}
\\
\mbox{}\hspace*{-2em}\mbox{Adaptation in discrete and continuous input spaces}
}
\vspace*{-4.5mm}
It is illustrative to consider a lowdimensional 
example how the Self-Organizing Map 
adapts to the structure of the stimuli data,
which is defined only by the input probability density.
Now the neural layer is chosen to be only one-dimensional,
and the first and last neuron are connected 
by periodic boundary conditions.

If the probability density is given by 
a finite sum of delta-peaks, 
and - provided a siutable parameter decay -
the neural weights of the Self-Organizing Map will converge
to these stimuli, 
as shown in Fig.~2 
and it will approximately 
find the shortest route visiting all stimuli
\cite{rittermartinetzschulten}.

\epsfig{file=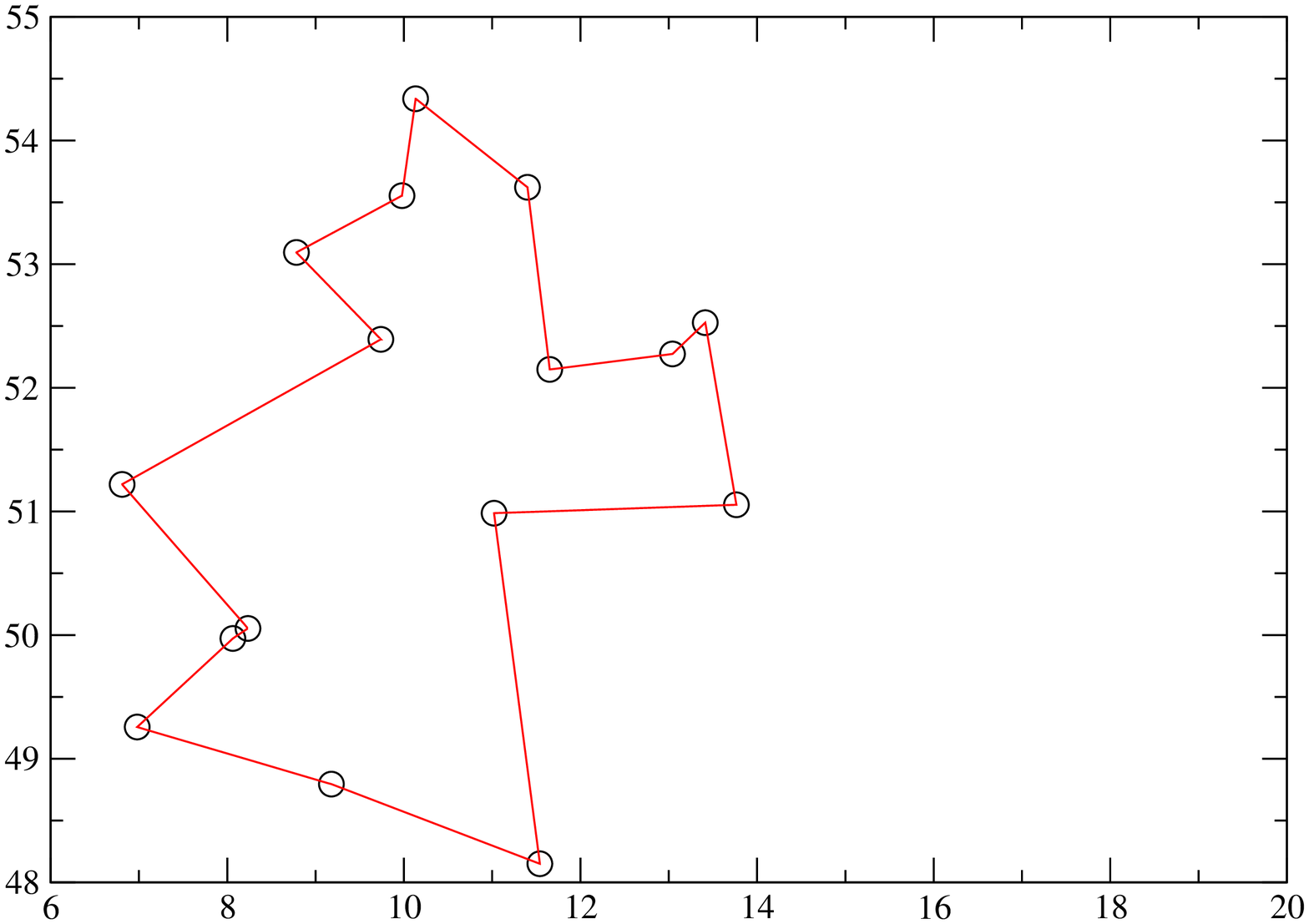,width=0.45\textwidth,angle=0}
\vspace*{.010cm}
\mbox{} \\ \noindent 
{\sl 
{\bf Fig. 2:} Solving the Travelling Salesman Problem
with the Kohonen Map.
}

If we call the stimuli `cities', we recognize 
this as the famous Travelling Salesman problem,
which is believed to be a NP-complete problem
- no algorithm can be found that computes the
optimal solution within a computation time
that scales polynomial with the number of
cities: Instead computation time scales exponentially.

Other algorithms have been given
that also give near-optimal solutions:
the thermodynamic-motivated 
 Simulated Annealing of Kirkpatrick \cite{kirkpatrick},
the neural approach by Hopfield and Tank \cite{hopfieldtank},
and the Elastic Net of Durbin and Willshaw \cite{durbinwillshaw}.
The latter two have been found to be limiting cases
of a unified approach \cite{simic90}.

If we now replace the input probability density
by a continuous one that may now be constant
within the country, and zero outside, as shown 
as background in 
 Fig.~3
the weight vectors will try to cover the 
countryside as good as possible, 
being in conflict between preserving 
topology as far as possible, 
and minimizing the reconstruction
error.
The necessary dimension reduction takes place
by a snake-like folding of the weights to 
locally step up and down in the excess dimensions.
For the input coming from the retina, this dimension reduction 
task is heavier (from 5 to 2 dimensions) \cite{obermayer92}.

\epsfig{file=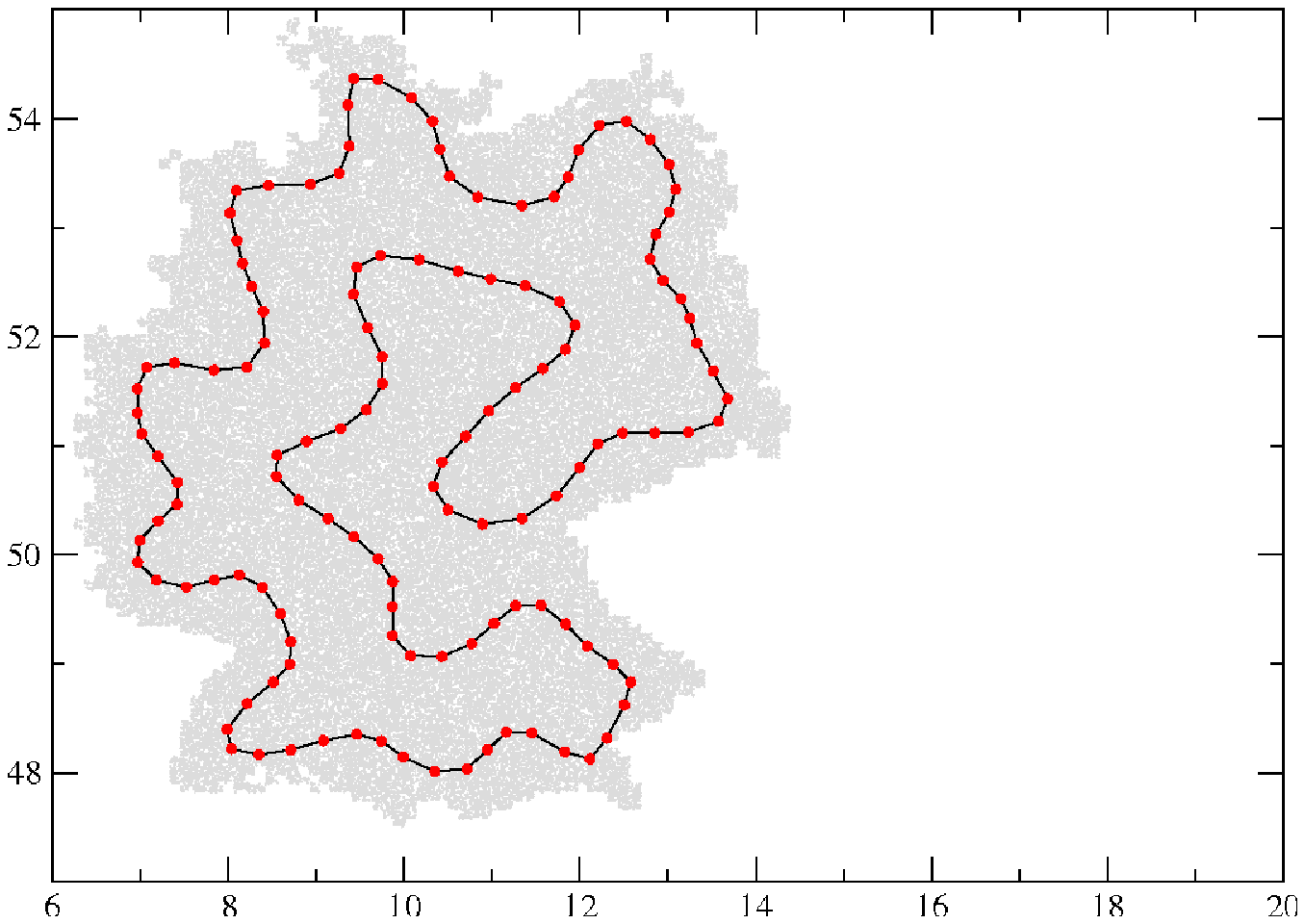,width=0.45\textwidth,angle=0}
\vspace*{.010cm} 
\mbox{} \\ \noindent 
{\sl 
{\bf Fig. 3:} For continuous input spaces of higher dimension, 
the Self-Organizing map approximates by 
maeandring structures.
}

\section{\large The Magnification Factor}

Depending on the (now assumed to be continuos) 
input probability density $P(\vec{v})$ of
the stimuli, an adaptive map algorithm can spend more neurons to
represent areas of higher probability density, according to a higher
resolution.

The magnification factor is defined as the density of
neurons $\vec{r}$ (i.~e. the density of
synaptic weight vectors $\vec{ w}_{\vec{r}}$)
per unit volume of input space, and therefore is given by the
inverse Jacobian of the mapping
from input space to neuron layer:
$M=|J|^{-1}=|\det({{d}}\vec{ w}/{{d}}\vec{r})|^{-1}$.
(In the following we
consider the onedimensional case of noninverting mappings, where
$J$ is positive.)
The magnification factor is a property of the networks' response
to a given probability density of stimuli $P(\vec{v})$.
To evaluate $M$ in higher dimensions, one in general has to
compute the  equilibrium state of the whole network and
needs therefore the complete global knowledge on $P(\vec{v})$.

For one-dimensional mappings (and possibly for special geometric
cases in higher dimensions) the magnification factor can follow
an universal magnification law, that is,
$M(\bar{\vec{ w}}(\vec{r}))$ is a function only of the local
probability density $P$ and independent of both the location
$\vec{r}$ in the neural layer and the location
$\bar{\vec{ w}}(\vec{r})$ in input space.

An optimal map from the view of information theory would
reproduce the input probability exactly
($M\sim P(\vec{v})^{\rho}$ with $\rho=1$), according
to a power law with exponent~1. This is equivalent to the
condition that all neurons in the layer are firing with same
probability. 
An exponent  \mbox{$\rho=0$}, on the other hand, corresponds to
a uniform distribution of weight vectors, which means there is no
adaptation to the stimuli at all.
So the magnification exponent is a direct indicator, how far a
Self Organizing Map algorithm is away from the optimum predicted
by information theory.

As the brain is assumed to be optimized by evolution 
for information procession, one would postulate that 
maximal mutual information is a sound principle
governing the setup of neural structures.
Such an algorithm of maximal mutual information
has been defined by Linsker \cite{linsker89}
using the gradient descend in mutual information.
It requires computationally costly integrations,
and has no local or other learning rule
that allows for biological motivation.

However, both biological network structures and technical
applications are (due to realization constraints) not necessarily
capable of reaching this optimum,
being especially for the brain 
under discussion \cite{plumbley}.
Even if one had quantitative experimental 
measurements of the magnification behaviour, the
question from what self-organizing dynamics 
the neural structure emerged remains.
So overall it is desirable to formulate other learning rules
that minimize mutual information in a simpler way.

To start with the simplest algorithm:
For the classical Kohonen algorithm the magnification law
(for onedimensional mappings) is
given by a power law $M(\bar{\vec{ w}}(\vec{r}))
\propto P(\bar{\vec{ w}}(\vec{r}))^{\rho}$ with exponent
$\rho=2/3$ \cite{ritter86}.
Although for a discrete neural layer and especially 
for neighborhood kernels with different shape and 
range there are corrections to the magnification law
\cite{rittermartinetzschulten,ritter91,dersch},
we consider the limit of a continuous neural layer,
and restrict our analysis to the onedimensional case.

\section{\large The Generalized Winner Relaxing Kohonen Algorithm}
We now consider an energy function that was at first
proposed in \cite{rittermartinetzschulten} 
for the classical Kohonen Algorithm,
and is given by the mean squared reconstruction error
of the resulting map.
For continuous input spaces, or when the borders of 
the voronoi tesselation shift across a localized stimulus,
there have to be corrections.

Now one can, if an energy function is desired, turn
the argumentation around: If the Self-Organizing Map 
has no energy function, and if 
the sqared reconstraction error
is an approximate one, start from this energy formula
and try to derive what learning rule will result.

Kohonen has,
utilizing some approximations,
shown in \cite{kohonen91} 
for the one- or  two-dimensional case 
that a gradient descent in 
the mean squared reconstruction error
results
in a
slightly different learning rule 
only for the winning neuron,
due to that
also the borders of the voronoi tesselation
are shifting
if one evaluates the gradient
with respect to a weight vector.

As the additional learning term
 implies an additional elastic relaxation for the
winning neuron, it is straightforward to call it
`Winner Relaxing' (WR) Kohonen algorithm.
As the relaxing term acts only in one direction,
the winner is relaxed to its neighbours, but the neighbours
stay unattracted, 
it can not strictly be interpreted as an 
elastic force or physical interaction.

It is straightforward to  generalize
the Winner Relaxing algorithm 
by introducing the free
parameter $\lambda$ to
the generalized Winner Relaxing Kohonen map 
\cite{claussendipl}
\begin{eqnarray}
\delta \vec{ w}_{\vec{r}} = \eta\{(\vec{v}-\vec{ w}_{\vec{r}})
                 g^{\gamma}_{\vec{r}\vec{s}}
   - \lambda   \delta_{\vec{r}\vec{s}}
   \sum_{\vec{r}^{'}\neq\vec{s}}
      g^{\gamma}_{\vec{r}^{'}\vec{s}}
      (\vec{v}-\vec{ w}_{\vec{r}^{'}})\},
\label{eq:wr_upd}
\end{eqnarray}
where $\vec{s}$ is the center of excitation for incoming stimulus
$\vec{v}$, and $g^{\gamma}_{\vec{r}\vec{s}}$ is a Gaussian function
of distance in the neural layer with characteristic length $\gamma$.
The original Algorithm proposed by Kohonen \cite{kohonen91} is
obtained for $\lambda=+1/2$, whereas the classical
Self Organizing Map Algorithm is obtained
for $\lambda=0$.
(Note that only for  $\lambda=+1/2$ the 
algorithm is associated with the potential function!)


\section{\large Magnification Exponent of the Generalized Winner-Relaxing Kohonen Algorithm
}

The necessary condition for the final state
of the algorithm  is that for all neurons
the expectation value of the learning step vanishes.
This gives a Chapman-Kolmogorov-Equation for 
the stochastic learning process of serial presentation.
This can be used also to derive the Magnification Law of the
the  Winner-Relaxing
Kohonen algorithm \cite{claussendipl}:
Insertion of the update rule 
into the stationarity condition
and integration
yields for
$(\bar{P}:=P(\bar{w}(r)))$
the differential equation
\begin{eqnarray}
0 = \gamma^2 (J \frac{{{d}}(\bar{P}J)}{{{d}}r}
  + \frac{\bar{P}J}{2} \frac{{{d}}J}{{{d}}r}
  + \lambda \frac{\bar{P}J}{2} \frac{{{d}}J}{{{d}}r}).
\end{eqnarray}
For $\gamma\neq 0,$    $P\neq 0,$
${{d}}\bar{P}/{{d}}r\neq 0$ and making the ansatz
$J(r)=J(\bar{P}(r))$ of an universal local magnification law
(that may be expected for the one-dimensional case)
we obtain the differential equation
\begin{eqnarray}
\frac{{{d}}J}{{{d}}\bar{P}}
&=& - \frac{2}{3+\lambda} \frac{J}{\bar{P}}
\end{eqnarray}
with its solution
(provided that $\lambda\neq{-3}$)
\cite{claussendipl}:
\begin{eqnarray}
M=\frac{1}{J} \sim P(v)^{\mbox{${\frac{2}{3+\lambda}}$}}.
\end{eqnarray}

For the Winner-Relaxing Kohonen Algorithm $(\lambda=1/2)$
the magnification factor follows an exact power law with
magnification exponent $\rho=4/7$, which is smaller than
$(\rho=2/3)$ for the classical Self Organizing Feature Map.
Although the Winner-Relaxing Kohonen Algorithm
is `somewhat faster' \cite{kohonen91} in the initial
ordering process, the resulting invariant mapping is
slightly less optimal in terms of information theory.

From this result one would try to invert the
Relaxing Effect by choice of negative
values for $\lambda$. The choice of $\lambda=-1$ would
lead to the magnification exponent one, if the algorithm
is stable for this parameter choice. 
This is tested by our numerical experiment described below.

\section{\large Numerical Verification of the
Magnification Law of the Winner Enforcing and Relaxing
Algorithms}
\vspace*{-2ex}
We used the following numerical setup: 
The network should map the unit interval to a 
onedimensional neural chain of 100 neurons.
The learning rate was $0.1$.
The stimulus probability density
was chosen exponentially as $\exp(-\beta{}w)$ with $\beta=4$.
After an adaptation process of $5\cdot{}10^7$ (Elastic Net)
 further $10\%$ of learning steps
were used to calculate average slope and its fluctuation
of $\log{}J$ as a function of $\log{}P.$ (The first and last $10\%$
of neurons were excluded to eliminate boundary effects).
The results are shown for several parameters in 
Fig. 4.

\epsfig{file=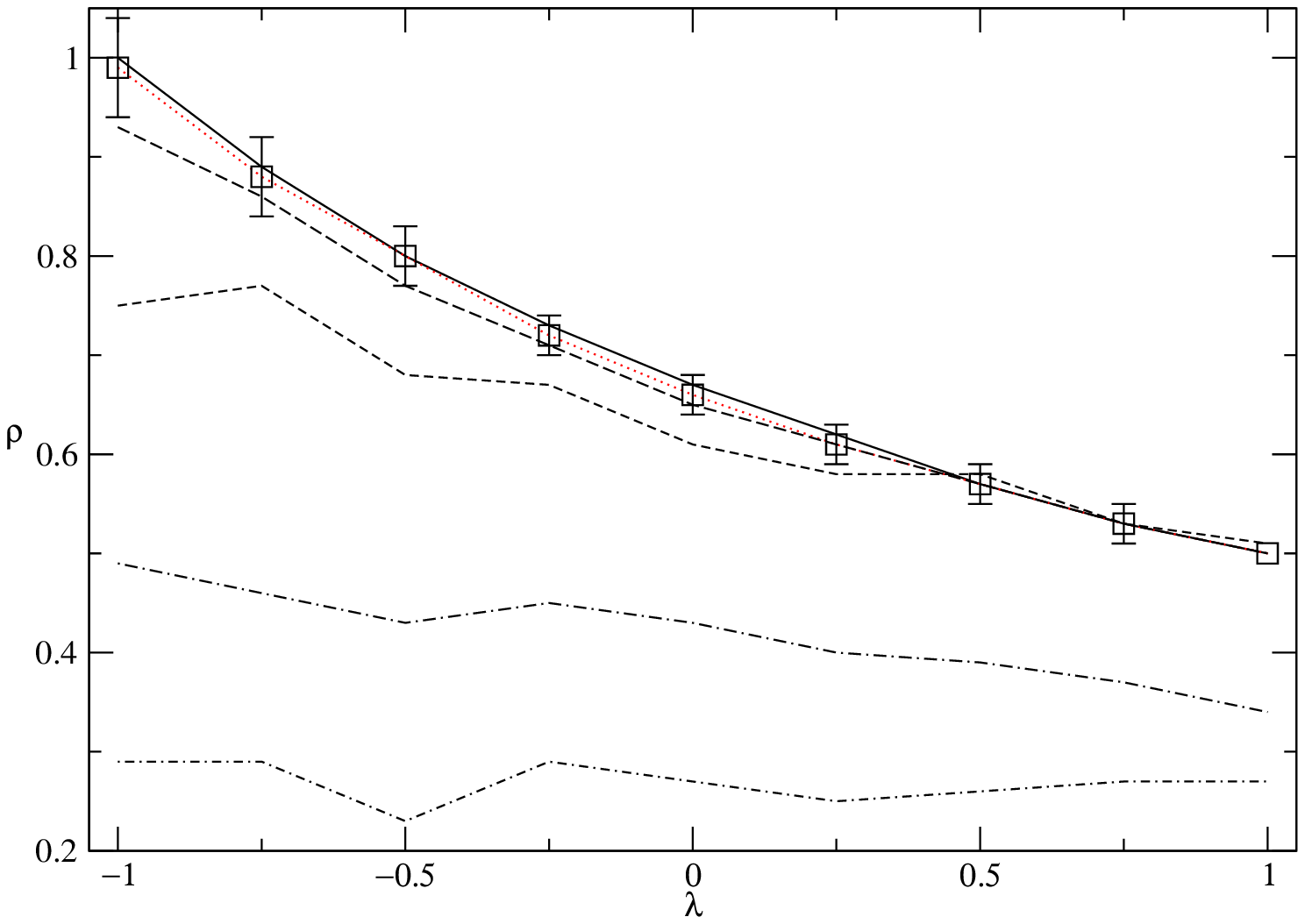,width=0.45\textwidth,angle=0}
\mbox{}\\\noindent
{\sl
{\bf Fig. 4:} 
Numerical Verification of the WRK/WEK magnification law.
The upper line is the thoretical prediction,
followed by $\gamma=5.0$ where the error bars show
that the theoretical prediction is met within the precision.
The lower lines are for $\gamma$ =2.0, 1.0, 0.5, and 0.1,
showing that a neighborhood interaction of system size
destroys adaptation.
}

For the Winner Relaxing and Enforcing Algorithm Family 
we did simulations for different $\lambda$ and 
neighborhood length $\gamma$. For small $\gamma$, the 
neighborhood interaction becomes too weak. If the Gaussian 
neighborhood extends over some neurons $\gamma=2$, $\gamma=5$,
the exponent follows the predicted dependence of $\lambda$
given by $2/(3+\lambda)$. For $|\lambda|>1$ we found the system
to become instable, this is the case where the additional update
term of the winner is larger than the sum over all other update
terms in the whole network.
The algorithm remains stable on both
stability borders $\lambda=+1$ and $\lambda=-1$,
and the escape time diverges approaching these boundaries
from outside. 
A detailed discussion of the numerical study will be
found in \cite{condmat0208414}.
As the relaxing effect of $\lambda>0$ is inverted for
$\lambda<0$, fluctuations are larger than in the Kohonen 
case.

Apart from the fact that the exponent can be varied by 
{\it a~priori} parameter choice between $1/2$ and $1$, the
simulations show that our Winner Enforcing Algorithm
is in fact able to establish information-theoretically
optimal self-organizing maps.

\section{\large Ordering behaviour of the WR and WE Kohonen maps}
\vspace*{-2ex}
One might suspect that the inversion of a smoothing term might
lead to larger fluctuations that could enlarge the time
needed for convergence.
Here the 
ordering behaviour is analyzed for a 
standard setup of 100 neurons with weights and
stimuli uniform in the unit interval,
with a high learning rate 
$\eta=1$ corresponding to parameters
used in the initial ordering phase (at that high
learning rate, out of the shown interval
the ordering time increased by magnitudes of order).

Fig.~5 
shows the averaged number of learning steps per neuron that
were needed until a monotonously increasing or
decreasing list of weights was reached.
In contrast to the initial exspectation, the
minimal ordering time is found neither for $\lambda=0$
(Self-Organizing Map) nor for the
energy-function associated Winner-Relaxing ($\lambda=1/2$) 
case.
Here we have the astonishing result that 
quicker ordering is not in complete contradiction to
near-infomax mapping and can both be 
realized with the Winner-Enhancing Kohonen Algorithm.

\epsfig{file=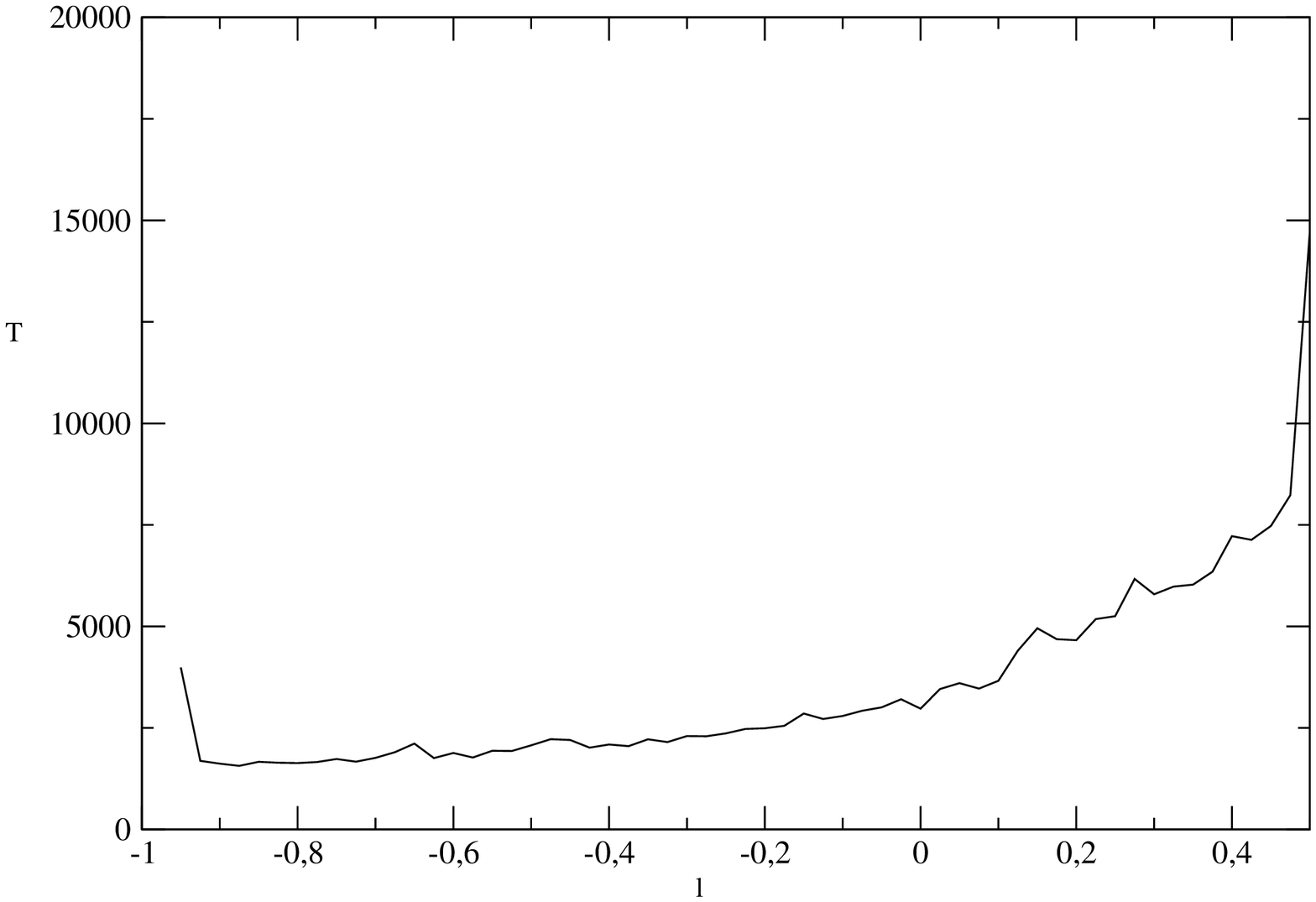,width=0.45\textwidth,angle=0}
\mbox{}\\ \noindent
{\sl
{\bf Fig. 5:} 
Ordering Behavior: Number of learning steps per neuron
as a function of parameter $\lambda$.
}
\vspace*{-3mm}

\section{\large Other recent approaches}
\vspace*{-3mm}  
After our first study\cite{claussendipl}, 
Herrmann et.al.
\cite{hermann}
introduced annother modification of the
learning process, which was also applied 
 to the Neural Gas (which is equivalent to
the Kohonen map without neighbour interaction)
\cite{villmann}.
This approach uses the central idea to 
make the learning rate $\eta$ locally dependent on the
input probability density by a power law
with an exponent that is related to the
desired magnification exponent, and also an exponent
1 can be obtained. 
As the input probability density should not be available
to the neuronal map that self-organizes 
from the stimuli drawn from that distribution,
it is estimated from the actual local reconstruction
mismatch (being an estimate for the size of the
voronoi cell) and from the time elapsed since
the last time being the winner. 
Due to this estimating character, the learning rate
has to be bounded in practical use.
From the computational point of view, 
one has to keep track of the time difference
between the firing of two neurons, which
introduces some memory term that needs extra storage,
and the local learning rate has to be computed,
which seems to be more costly than 
the Winner Enhancing Kohonen.

Other modifications consider the {\sl selection} of the winner
to be probabilistic, leading to much more  elegant
statistical approaches  to  potential functions
(see 
Graepel et al. \cite{graepel97}
and Heskes
\cite{heskes99energyfunctions}).

The robustness of the Winner-Relaxing principle
has been demonstrated by transferring it to the 
Neural Gas architecture,
which now allows by this simple approach
to obtain a magnification exponent 1
in arbitrary dimension
\cite{clauvillesann03}.
\vspace*{-3mm}


\section{\large Conclusions}
\vspace*{-3mm}
Feature maps are self-organizing structures 
in the brain 
and in computational neuroscience
that can efficiently represent high-dimensional
and complex input spaces.
Retina, skin and other perceptual receptor areas are represented
in a topology-preserving manner, i.~e. if adjacent neighbours are firing,
the active receptor cells also are adjacent. 
The detailed structure of these neural maps, including all synaptic
connections, cannot be coded genetically, so it appears necessary
to develop models that set up their structure by a self-organizing
progress. 
The Self-Organizing Map has become the most prominent
model and been applied to many technical problems. 
Several other models, the Linsker   
Algorithm, the Elastic Net 
Algorithm and the Winner Relaxing Kohonen Algorithm have also 
been considered as models for feature maps and used in technical
applications.
Most of them 
follow from an extremal principle, given by information theory,
physical motivations, or reconstruction error.
But what extremal principles govern the feature maps in the 
brain? 

To answer this question finally would reqire 
more quantitative data about the magnification
behaviour in experiments, 
which then would give a basis to judge how close
nature comes to the optimum given by information theory.

Magnification-adjustable models as the Winner-Relaxing and
Winner-Enhancing Self-Organizing map can become 
a valuable tool for comparison with experiments
and further refinement of the theoretical understanding
of the brain.
\vspace*{-4.5mm}


\end{multicols}

\end{document}